\documentclass[aps,preprint,nofootinbib,eqsecnum]{revtex4}%
\usepackage{amsmath}
\usepackage{amsfonts}
\usepackage{amssymb}
\usepackage{graphicx}
\usepackage{amscd}%
\providecommand{\U}[1]{\protect\rule{.1in}{.1in}}

\allowdisplaybreaks
\begin{document}
\leftline {USC-08/HEP-B2 \hfill 0804.1585 [hep-th]}{}{\vskip-1cm}

{\vskip2cm}

\begin{center}
{\Large \textbf{Gravity in 2T-Physics}}\footnote{This work was partially
supported by the US Department of Energy under grant number
DE-FG03-84ER40168.}{\Large \textbf{\ }}

{\vskip0.8cm}

\textbf{Itzhak Bars}

{\vskip0.8cm}

\textsl{Department of Physics and Astronomy}

\textsl{University of Southern California,\ Los Angeles, CA 90089-2535 USA}

{\vskip1.5cm} \textbf{Abstract}
\end{center}

The field theoretic action for gravitational interactions in $d+2$ dimensions
is constructed in the formalism of 2T-physics. General Relativity in $d$
dimensions emerges as a shadow of this theory with one less time and one less
space dimensions. The gravitational constant turns out to be a shadow of a
dilaton field in $d+2$ dimensions that appears as a constant to observers
stuck in $d$ dimensions. If elementary scalar fields play a role in the
fundamental theory (such as Higgs fields in the Standard Model coupled to
gravity), then their shadows in $d$ dimensions must necessarily be
\textit{conformal} scalars. This has the physical consequence that the
gravitational constant changes at each phase transition (inflation, grand
unification, electro-weak, etc.), implying interesting new scenarios in
cosmological applications. The fundamental action for pure gravity, which
includes the spacetime metric $G_{MN}\left(  X\right)  ,$ the dilaton
$\Omega\left(  X\right)  $ and an additional auxiliary scalar field $W\left(
X\right)  ,$ all in $d+2$ dimensions with two times, has a mix of gauge
symmetries to produce appropriate constraints that remove all ghosts or
redundant degrees of freedom. The action produces on-shell classical field
equations of motion in $d+2$ dimensions, with enough constraints for the
theory to be in agreement with classical General Relativity in $d$ dimensions.
Therefore this action describes the correct classical gravitational physics
directly in $d+2$ dimensions. Taken together with previous similar work on the
Standard Model of particles and forces, the present paper shows that
2T-physics is a general consistent framework for a physical theory.
Furthermore, the 2T-physics approach reveals more physical information for
observers stuck in the shadow in $d$ dimensions in the form of hidden
symmetries and dualities, that is largely concealed in the usual one-time
formulation of physics.

\newpage

\section{Gravitational background fields in 2T-physics}

Previous discussions on gravitational interactions in the context of
2T-physics appeared in \cite{2tbacgrounds}\cite{2tfield}\cite{2treviews}.
There it was shown how to formulate the motion of a particle in background
fields (including gravity, electromagnetism, high spin fields) with a target
spacetime in $d+2$ dimensions with two times. The previous approach was a
worldline formalism in which consistency with an Sp$\left(  2,R\right)  $
gauge symmetry produced some constraints on the backgrounds. Those
restrictions should be regarded as gauge symmetry \textit{kinematical
}constraints on the background fields, which can be used to eliminate ghosts
and redundant degrees of freedom by choosing a unitary gauge if one wishes to
do so. Consistent with the notion of backgrounds, the Sp$\left(  2,R\right)  $
constraints by themselves did not impose any conditions on the
\textit{dynamics} of the physical background fields that survive after
choosing a unitary gauge.

In the present paper we construct the \textit{off-shell} field theoretic
action for Gravity in $d+2$ dimensions, that not only reproduces the correct
Sp$\left(  2,R\right)  $ gauge symmetry kinematical constraints mentioned
above when the fields are on-shell, but also yields the on-shell or off-shell
\textit{dynamics} of gravitational interactions. This $d+2$ formulation of
gravity is in full agreement with classical General Relativity in $\left(
d-1\right)  +1$ dimensions with one time as described in the Abstract.

We will use the brief notation GR$_{d}$ to refer to the emergent form of
General Relativity, which is usual GR with some additional constraints that
are explained below, while the notation GR$_{d+2}$ is reserved for the parent
theory from which GR$_{d}$ is derived by solving the kinematic constraints. So
GR$_{d}$ can be regarded as a lower dimensional holographic shadow of
GR$_{d+2}$ that captures the gauge invariant physical sector that satisfies
the Sp$\left(  2,R\right)  $ kinematic constraints. There are however other
holographic shadows of the same GR$_{d+2}$ that need not look like GR$_{d}$
but are related to it by duality transformations. These shadows, and the
relations among them, provide additional information about the nature of
gravity that is not captured by the usual one-time formulation of physics.

The key element of 2T-physics is a worldline Sp$\left(  2,R\right)  $ gauge
symmetry which acts in \textit{phase space} and makes position and momentum
$\left(  X^{M}\left(  \tau\right)  ,P_{M}\left(  \tau\right)  \right)  $
indistinguishable at any worldline instant $\tau$ \cite{2treviews}. This
Sp$\left(  2,R\right)  $ gauge symmetry is an upgrade of worldline $\tau$
reparametrization to a higher gauge symmetry. It cannot be realized if target
spacetime has only one time dimension. It yields nontrivial physical content
only if the target spacetime $X^{M}$ includes two time dimensions.
Simultaneously, this larger worldline gauge symmetry plays a crucial role to
remove all unphysical degrees of freedom in a 2T spacetime, just as worldline
reparametrization removes unphysical degrees of freedom in a 1T spacetime.
Furthermore, more than two times cannot be permitted because the Sp$\left(
2,R\right)  $ gauge symmetry cannot remove the ghosts of more than 2 timelike dimensions.

We could discuss the field theory for Gravity directly, but it is useful to
recall some aspects of the worldline Sp$\left(  2,R\right)  $ formalism that
motivates this construction. The general 2T-physics worldline action for a
spin zero particle moving in any background field is given by
\cite{2tbacgrounds}%
\begin{equation}
S=\int d\tau(\partial_{\tau}X^{M}P_{M}\left(  \tau\right)  -\frac{1}{2}%
A^{ij}\left(  \tau\right)  ~Q_{ij}\left(  X\left(  \tau\right)  ,P\left(
\tau\right)  \right)  \,). \label{actionW}%
\end{equation}
This action has local Sp$\left(  2,R\right)  $ symmetry on the worldline
\cite{2tbacgrounds}. The 3 generators of Sp$\left(  2,R\right)  $ are
described by the symmetric tensor $Q_{ij}=Q_{ji}$ with $i=1,2,$ and the gauge
field is $A^{ij}\left(  \tau\right)  .$ The background fields as functions of
spacetime $X^{M}$ are the coefficients in the expansion of $Q_{ij}\left(
X,P\right)  $ in powers of momentum, $Q_{ij}\left(  X,P\right)  =Q_{ij}%
^{0}\left(  X\right)  +Q_{ij}^{M}\left(  X\right)  P_{M}+Q_{ij}^{MN}\left(
X\right)  P_{M}P_{N}+\cdots.$

In the current paper we wish to describe only the gravitational background.
Therefore, specializing to a simplified version of \cite{2tbacgrounds} we take
just the following form of $Q_{ij}\left(  X,P\right)  $%
\begin{equation}
Q_{11}=W\left(  X\right)  ,\quad Q_{12}=V^{M}\left(  X\right)  P_{M},\quad
Q_{22}=G^{MN}\left(  X\right)  P_{M}P_{N},\label{q2}%
\end{equation}
which includes the gravitational metric $G^{MN}\left(  X\right)  ,$ together
with an auxiliary scalar field $W\left(  X\right)  $ and a vector field
$V^{M}\left(  X\right)  .$ A basic requirement for the Sp$\left(  2,R\right)
$ gauge symmetry of the worldline action is that the generators $Q_{ij}\left(
X,P\right)  $ must satisfy the Sp$\left(  2,R\right)  $ Lie algebra under
Poisson brackets. This requirement turns into certain kinematical constraints
on the background fields $\left(  W\left(  X\right)  ,V^{M}\left(  X\right)
,G^{MN}\left(  X\right)  \right)  ,$ which are obtained by demanding closure
of Sp$\left(  2,R\right)  $ under Poisson brackets $\left\{  A,B\right\}
\equiv\frac{\partial A}{\partial X^{M}}\frac{\partial B}{\partial P_{M}}%
+\frac{\partial A}{\partial P_{M}}\frac{\partial B}{\partial X^{M}}$ as
follows \cite{2tbacgrounds}\cite{2tfield}%
\begin{align}
\left\{  Q_{11},Q_{22}\right\}   &  =4Q_{12}\;\rightarrow V^{M}=\frac{1}%
{2}G^{MN}\partial_{N}W,\label{sp1}\\
\left\{  Q_{11},Q_{12}\right\}   &  =2Q_{11}\;\rightarrow\;V^{M}\partial
_{M}W=2W,\;\label{sp2}\\
\left\{  Q_{22},Q_{12}\right\}   &  =-2Q_{22}\;\rightarrow\pounds _{V}%
G^{MN}=-2G^{MN}.\label{sp3}%
\end{align}
In the last line $\pounds _{V}G^{MN}$ is the Lie derivative of the metric,
which is a general coordinate transformation of the metric using the vector
$V^{M}\left(  X\right)  $ as the parameter of transformation%
\begin{align}
-2G^{MN} &  =V^{K}\partial_{K}G^{MN}-\partial_{K}V^{M}G^{KN}-\partial_{K}%
V^{N}G^{MK}\label{sp4}\\
&  =-\nabla^{M}V^{N}-\nabla^{N}V^{M}\equiv\pounds _{V}G^{MN}\label{sp45}%
\end{align}
The equivalence of the expressions in (\ref{sp4},\ref{sp45}) is seen by
replacing every derivative in (\ref{sp4}) by covariant derivatives using the
Christoffel connection $\Gamma_{MN}^{P}$, such as $\nabla_{P}V^{N}%
=\partial_{P}V^{N}+\Gamma_{PQ}^{N}V^{Q},$ and recalling that the covariant
derivative of the metric vanishes $\nabla_{K}G^{MN}=0:$%
\begin{equation}
\nabla_{K}G^{MN}=0\;\leftrightarrow\Gamma_{MN}^{P}=\frac{1}{2}G^{PQ}\left(
-\partial_{Q}G_{MN}+\partial_{M}G_{NQ}+\partial_{N}G_{MQ}\right)
.\label{christoffel}%
\end{equation}
We can deduce that the above relations imply that $G_{MN}$ can be written as
\begin{equation}
G_{MN}=\nabla_{M}V_{N}=\frac{1}{2}\nabla_{M}\partial_{N}W.\label{sp5}%
\end{equation}
This is proven by inserting the expression for the Christoffel connection in
$G_{MN}=\nabla_{M}V_{N}=\partial_{M}V_{N}-\Gamma_{MN}^{P}V_{P}$ and using
(\ref{sp1}-\ref{sp4}).

There are an infinite number of solutions \cite{2tbacgrounds} that satisfy
(\ref{sp1}-\ref{sp5}). An example is flat spacetime
\begin{equation}
W_{flat}\left(  X\right)  =X\cdot X,\;\;V_{flat}^{M}\left(  X\right)
=X^{M},\;\;G_{flat}^{MN}\left(  X\right)  =\eta^{MN}. \label{flat}%
\end{equation}
This satisfies the Sp(2,R) relations (\ref{sp1}-\ref{sp5}). In this case the
Sp$\left(  2,R\right)  $ generators are simply
\begin{equation}
Q_{11}^{flat}=X\cdot X,\quad Q_{12}^{flat}=X\cdot P,\quad Q_{22}^{flat}=P\cdot
P. \label{flatQ}%
\end{equation}
This flat background has an SO$\left(  d,2\right)  $ global symmetry (Killing
vectors of the flat metric $\eta_{MN}$) whose generators $L^{MN}=X^{M}%
P^{N}-X^{N}P^{M}$ commute with the dot products in (\ref{flatQ})$.$

The phase space $\left(  X^{M},P_{M}\right)  $ and the background fields
$W\left(  X\right)  ,V^{M}\left(  X\right)  ,G^{MN}\left(  X\right)  $ are
restricted by the Sp(2,R) relations (\ref{sp1}-\ref{sp5}) as well as by the
requirement of Sp$\left(  2,R\right)  $ gauge invariance $Q_{ij}\left(
X,P\right)  =0$ in the physical subspace. The latter is derived from the
action (\ref{actionW}) as the equation of motion for the gauge field $A^{ij}.$
This combination of constraints are just the right amount to remove ghosts
from a 2T spacetime and end up with a shadow sub-phase-space $\left(  x^{\mu
},p_{\mu}\right)  $ with a 1T spacetime which describes the gauge fixed
physical sector. There are no non-trivial solutions if the higher spacetime
has fewer than 2 timelike dimensions. This is easy to verify for the flat
example (\ref{flat}). Furthermore, if the higher spacetime has more than 2
timelike dimensions there are always ghosts. Hence the Sp$\left(  2,R\right)
$ gauge symmetry demands precisely 2 timelike dimensions, no less and no
more\footnote{A more general argument that applies to all backgrounds is the
following. By canonical transformations that do not change the signature, the
first two constraints $Q_{11},Q_{12}$ can always be brought to the flat form,
while $Q_{22}$ has the backgrounds (second reference in \cite{2tbacgrounds}).
Then non-trivial solutions require 2 times. Another point is that the
signature of the Sp$\left(  2,R\right)  $ parameters, which is the same as
SO$\left(  1,2\right)  $ with 1 space and two times, determines the signature
of the constraints and of the removable degrees of freedom from $\left(
X^{M},P_{M}\right)  $.}.

The solution of (\ref{sp1}-\ref{sp5}) at the classical level was obtained in
\cite{2tbacgrounds}\cite{2tfield}, where it was shown that the worldline
action (\ref{actionW}) reduces (as one of the shadows) to the well known
1-time worldline action of a particle moving in an arbitrary gravitational
background field $g_{\mu\nu}\left(  x^{\mu}\right)  $ in $d$ dimensions%
\begin{equation}
S=\int d\tau(\partial_{\tau}x^{\mu}p_{\mu}\left(  \tau\right)  -\frac{1}%
{2}A^{22}\left(  \tau\right)  ~g^{\mu\nu}\left(  x\left(  \tau\right)
\right)  p_{\mu}\left(  \tau\right)  p_{\nu}\left(  \tau\right)  \,).
\label{worldlineGravity}%
\end{equation}
This 1T action has enough well known gauge symmetry to remove ghosts in
1T-physics. This remaining gauge symmetry is part of the original Sp$\left(
2,R\right)  .$

This fixing of gauges to a unitary gauge, demonstrates that the Sp(2,R)
relations (\ref{sp1}-\ref{sp5}) have the right amount of gauge symmetry to
remove ghosts. Hence the 2T-physics approach provides a physical theory for
gravity formulated directly in the higher spacetime $X^{M}$ in $d+2$
dimensions with two times in the form of the action (\ref{actionW}), as long
as the background fields $W\left(  X\right)  ,V^{M}\left(  X\right)
,G^{MN}\left(  X\right)  $ satisfy the Sp(2,R) kinematic constraints
(\ref{sp1}-\ref{sp5}) that are also formulated directly in $d+2$ dimensions.

Note however that the Sp(2,R) constraints are not enough to give the dynamical
equations that the gravitational metric $g^{\mu\nu}\left(  x\right)  $ in
$\left(  d-1\right)  +1$ dimensions should satisfy. To do this we must build a
field theoretic action in $d+2$ dimensions that not only gives correctly the
Sp(2,R) kinematic constraints (\ref{sp1}-\ref{sp5}), but also gives dynamical
equations in $d+2$ dimensions for the metric $G_{MN}\left(  X\right)  ,$ and
auxiliary fields $W\left(  X\right)  ,V^{M}\left(  X\right)  ,$ which in turn
correctly reproduce the equations of General relativity for the metric
$g_{\mu\nu}\left(  x\right)  .$ This is what we will present in the rest of
this paper.

\section{Gravitational Action \label{GravAction}}

The first kinematic equation (\ref{sp1}) will be imposed from the start, so
the auxiliary field $V^{M}\left(  X\right)  $ will not be included as a
fundamental one in the action, but instead will be replaced by $V_{M}=\frac
{1}{2}\partial_{M}W$ consistent with (\ref{sp1}). Recall that $Q_{11}=W\left(
X\right)  =0$ is one of the Sp$\left(  2,R\right)  $ constraints of the
worldline theory. To implement this constraint covariantly in $d+2$ dimensions
we follow the methods that were successful in flat space \cite{2tbrst2006}%
\cite{2tstandardM}, namely include a delta function as part of the volume
element $\delta\left(  W\left(  X\right)  \right)  d^{d+2}X$ in the definition
of the action of 2T field theory\footnote{Some studies for conformal gravity
in 4+2 dimensions using Dirac's approach to conformal symmetry \cite{Dirac}%
-\cite{marnelius2} also use fields in 4$+2$ dimensions and include a delta
function \cite{vasiliev}\cite{marnelius2} (see also \cite{marnelius}). Their
focus is conformal gravity aiming for and constructing a totally different
action. While we have some overlap of methods with \cite{vasiliev}%
\cite{marnelius2}, we have important differences right from the start. They
impose kinematic constraints as additional conditions that do not follow from
the action, as we did also in our older work \cite{2tfield}. These are related
to the conceptually more general Sp(2,R) constraints in 2T-physics. The new
progress in 2T field theory since \cite{2tbrst2006}\cite{2tstandardM} is to
derive the constraints as well the dynamics from the action, without imposing
them externally. In our present work the unusual piece of the action S$_{W},$
with $W$ a field varied like any other, are the new crucial ingredients in
curved space that allows us to derive all Sp$\left(  2,R\right)  $ constraints
from the action, and leads to the new physical consequences.}. The field $W$
will appear in other parts of the action as well. In flat space $W\left(
X\right)  $ was a fixed background $W_{flat}\left(  X\right)  =X\cdot X,$ but
in the present case it is a field that will be allowed to vary as any other.
In addition to $W\left(  X\right)  $ and $G_{MN}\left(  X\right)  $ we will
need also the dilaton field $\Omega\left(  X\right)  $ in order to impose
consistency with the kinematic constraints (\ref{sp1}-\ref{sp5}) required by
the underlying Sp$\left(  2,R\right)  .$ The dilaton plays a similar role even
in flat 2T field theory especially when $d\neq4~$\cite{2tstandardM}$.$ Our
proposed action for the 2T gravity triplet $G_{MN},\Omega,W$ is
\begin{align}
S  &  =S_{G}+S_{\Omega}+S_{W}\label{Stot}\\
S_{G}  &  \equiv\gamma\int d^{d+2}X~\delta\left(  W\right)  \sqrt{G}%
~\Omega^{2}R\left(  G\right) \label{SG}\\
S_{\Omega}  &  \equiv\gamma\int d^{d+2}X~\delta\left(  W\right)  \sqrt
{G}\left\{  \frac{1}{2a}\partial\Omega\cdot\partial\Omega-V\left(
\Omega\right)  \right\} \label{Sphi}\\
S_{W}  &  =\gamma\int d^{d+2}X~\delta^{\prime}\left(  W\right)  \sqrt
{G}\left\{  \Omega^{2}\left(  4-\nabla^{2}W\right)  +\partial W\cdot
\partial\Omega^{2}\right\}  \label{SW}%
\end{align}
Note that the last term in the action $S_{W}$ contains $\delta^{\prime}\left(
W\right)  $ rather than $\delta\left(  W\right)  .$ The overall constant
$\gamma$ is a volume renormalization constant that also appears in flat 2T
field theory (\cite{2tstandardM}\cite{emergentfieldth1}\cite{emergentfieldth2}%
), and is specified after Eq.(\ref{gamma}). Demanding consistency with the
Sp$\left(  2,R\right)  $ kinematic constraints (\ref{sp1}-\ref{sp5}) will fix
the constant $a$ uniquely to
\begin{equation}
a=\frac{\left(  d-2\right)  }{8\left(  d-1\right)  }. \label{a}%
\end{equation}
As will be explained below, for this special value of $a$, the
\textquotedblleft conformal shadow\textquotedblright\ in $d$ dimensions has an
accidental local Weyl symmetry (even though the $d+2$ theory does not have it).

The action above is a no scale theory. The dimensionful gravitational constant
will develop spontaneously from a vacuum expectation value of the dilaton
$\langle\Omega\rangle\neq0.$ The corresponding Goldstone boson as seen by
observers in $d$ dimensions is gauge freedom removable by the accidental Weyl
gauge symmetry.

The various factors in the action involving powers of $\Omega$ are determined
as follows. We assign engineering dimensions for $X^{M},G_{MN},\Omega,W,$
which are consistent with their flat counterparts in (\ref{flat}), as follows%
\begin{equation}
\dim\left(  X^{M}\right)  =1,\;\dim\left(  G_{MN}\right)  =0,\;\dim\left(
W\right)  =2,\;\dim\Omega=-\frac{d-2}{2}. \label{dims}%
\end{equation}
Accordingly, powers of the dilaton $\Omega$ are inserted as shown to insure
that the action is dimensionless $\dim\left(  S\right)  =0.$ The underlying
reason for this is a gauge symmetry, that we called the 2T gauge symmetry in
field theory \cite{2tstandardM}, which becomes valid when the factors of
$\Omega$ are included. The dimensions (\ref{dims}) will appear in the
Sp$\left(  2,R\right)  $ kinematic equations that follow from the action, and
coincide precisely with the kinematic constraints (\ref{sp2},\ref{sp3}) that
are required by the worldline Sp$\left(  2,R\right)  $ gauge symmetry$.$ These
turn into homogeneity constraints in flat space, when $V_{flat}^{M}=X^{M}$ and
$X\cdot\partial W_{flat}=2W_{flat}$ and $X\cdot\partial G_{flat}^{MN}=0,$
which are consistent with $\dim\left(  W\right)  =2,\dim\left(  G_{MN}\right)
=0$ respectively as given in (\ref{dims}). The consistency of the kinematic
equations with each other (equivalently the gauge symmetry) restricts the form
of self interactions of the scalar to the form%
\begin{equation}
V\left(  \Omega\right)  =\frac{\lambda\left(  d-2\right)  }{2d}\Omega
^{\frac{2d}{d-2}} \label{Vphi}%
\end{equation}
where the arbitrary constant $\lambda$ is dimensionless.

\section{Equations of motion for $G_{MN}$ \label{eomG}}

We first concentrate on $S_{G}.$ Using the variational formulas
\begin{equation}
\delta\sqrt{G}=-\frac{1}{2}\sqrt{G}G_{MN}\delta G^{MN},\;\delta R\left(
G\right)  =\left\{  R_{MN}+\left(  G_{MN}\nabla^{2}-\nabla_{M}\nabla
_{N}\right)  \right\}  \delta G^{MN}, \label{varG}%
\end{equation}
and doing integration by parts as needed, we obtain the following variation of
$S_{G}$ with respect to the metric%
\begin{align}
\delta_{G}\left(  S_{G}\right)   &  =\gamma\int d^{d+2}X~\delta\left(
W\right)  ~\Omega^{2}\delta_{G}\left(  \sqrt{G}R\left(  G\right)  \right)
=\gamma\int d^{d+2}X\sqrt{G}\delta G^{MN}\left(  V_{MN}^{G}\right) \\
V_{MN}^{G}  &  \equiv\delta\left(  W\right)  ~\Omega^{2}\left(  R_{MN}%
-\frac{1}{2}G_{MN}R\right)  +\left(  G_{MN}\nabla^{2}-\nabla_{M}\nabla
_{N}\right)  \left(  \delta\left(  W\right)  ~\Omega^{2}\right)  \label{vG}%
\end{align}
The last term will generate terms proportional to $\delta\left(  W\right)  ,$
$\delta^{\prime}\left(  W\right)  ,$ $\delta^{\prime\prime}\left(  W\right)  $
as follows%
\begin{align}
&  \left(  G_{MN}\nabla^{2}-\nabla_{M}\nabla_{N}\right)  \left(  \delta\left(
W\right)  ~\Omega^{2}\right) \nonumber\\
&  =\left\{
\begin{array}
[c]{l}%
\delta\left(  W\right)  \left[  G_{MN}\nabla^{2}\Omega^{2}-\nabla_{M}%
\partial_{N}\Omega^{2}\right] \\
+\delta^{\prime}\left(  W\right)  \left[
\begin{array}
[c]{c}%
2G_{MN}\partial W\cdot\partial\Omega^{2}-2\partial_{M}W\partial_{N}\Omega
^{2}\\
+\Omega^{2}\left(  G_{MN}\nabla^{2}W-\nabla_{M}\partial_{N}W\right)
\end{array}
\right] \\
+\delta^{\prime\prime}\left(  W\right)  \Omega^{2}\left[  G_{MN}\partial
W\cdot\partial W-\partial_{M}W\partial_{N}W\right]
\end{array}
\right\}  \label{vG2}%
\end{align}
Additional terms in the action are needed to modify the expressions
proportional to $\delta^{\prime}\left(  W\right)  ,\delta^{\prime\prime
}\left(  W\right)  $ because requiring $\delta_{G}\left(  S_{G}\right)  $ to
vanish on its own would put severe and inconsistent constraints on $G_{MN}$
and $\Omega$ that are incompatible with the Sp$\left(  2,R\right)  $ kinematic
conditions in (\ref{sp1}-\ref{sp5}). This is the first reason for introducing
the additional term $S_{W}$ which miraculously produces just the right
structure of variational terms that make the Sp$\left(  2,R\right)  $
constraints (\ref{sp1}-\ref{sp5}) compatible with the equations of motion
derived from the action. Actually $S_{W}$ performs a few more miracles
involving the variations of $\Omega$ and $W$ as well, as we will see below.

Thus let us study the variation of $S_{W}$ with respect to $\delta G^{MN}$%
\begin{equation}
\delta_{G}\left(  S_{W}\right)  =\gamma\int d^{d+2}X~~\delta^{\prime}\left(
W\right)  \left\{
\begin{array}
[c]{c}%
\left(  4\delta_{G}\sqrt{G}-\partial_{M}\left(  \delta_{G}\left(  \sqrt
{G}G^{MN}\right)  \partial_{N}W\right)  \right)  \Omega^{2}\\
+\delta_{G}\left(  \sqrt{G}G^{MN}\right)  \partial_{M}W\partial_{N}\Omega^{2}%
\end{array}
\right\}  .
\end{equation}
After an integration by parts this gives $\delta_{G}\left(  S_{W}\right)
=\gamma\int d^{d+2}X~\sqrt{G}\delta G^{MN}\left(  V_{MN}^{W}\right)  $ with
\begin{equation}
V_{MN}^{W}\equiv a\left\{
\begin{array}
[c]{l}%
+\delta^{\prime}\left(  W\right)  \left[  2\partial_{M}W\partial_{N}\Omega
^{2}-G_{MN}\left(  2\Omega^{2}+\partial W\cdot\partial\Omega^{2}\right)
\right] \\
+\delta^{\prime\prime}\left(  W\right)  \Omega^{2}\left[  \partial
_{M}W\partial_{N}W-\frac{1}{2}G_{MN}\partial W\cdot\partial W\right]
\end{array}
\right\}  . \label{vW}%
\end{equation}
We will also need the variation of $S_{\Omega}$ with respect to $\delta
G^{MN},$ but this contains only $\delta\left(  W\right)  $
\begin{align}
\delta_{G}\left(  S_{\Omega}\right)   &  =\gamma\int d^{d+2}X~\sqrt{G}\delta
G^{MN}\left(  V_{MN}^{\Omega}\right)  ,\\
V_{MN}^{\Omega}  &  \equiv\delta\left(  W\right)  \left[  \frac{1}{2a}%
\partial_{M}\Omega\partial_{N}\Omega+G_{MN}\left(  -\frac{1}{4a}\partial
\Omega\cdot\partial\Omega+\frac{1}{2}V\left(  \Omega\right)  \right)  \right]
. \label{vPhi}%
\end{align}
The vanishing of the total variation $\delta_{G}\left(  S_{G}+S_{W}+S_{\Omega
}\right)  =\gamma\int d^{d+2}X~\sqrt{G}\delta G^{MN}\left(  V_{MN}\right)  =0$
gives
\begin{align}
V_{MN}  &  =\delta\left(  W\right)  V_{MN}^{\left(  0\right)  }+\delta
^{\prime}\left(  W\right)  V_{MN}^{\left(  1\right)  }+\delta^{\prime\prime
}\left(  W\right)  V_{MN}^{\left(  2\right)  }=0,\label{Vtot}\\
V_{MN}^{\left(  0\right)  }  &  \equiv~\left[
\begin{array}
[c]{l}%
\Omega^{2}\left(  R_{MN}-\frac{1}{2}G_{MN}R\right)  +\left(  G_{MN}\nabla
^{2}\Omega^{2}-\nabla_{M}\partial_{N}\Omega^{2}\right) \\
\frac{1}{2a}\partial_{M}\Omega\partial_{N}\Omega+G_{MN}\left(  -\frac{1}%
{4a}\partial\Omega\cdot\partial\Omega+\frac{1}{2}V\left(  \Omega\right)
\right)
\end{array}
\right]  ,\label{Vtot1}\\
V_{MN}^{\left(  1\right)  }  &  \equiv\Omega^{2}\left[  G_{MN}\left(
-6+\nabla^{2}W+\partial W\cdot\partial\ln\Omega^{2}\right)  -\nabla
_{M}\partial_{N}W\right]  ,\label{Vtot2}\\
V_{MN}^{\left(  2\right)  }  &  \equiv\frac{1}{2}\Omega^{2}G_{MN}\left(
\partial W\cdot\partial W-4W\right)  . \label{Vtot3}%
\end{align}
The vanishing expression $\frac{1}{2}\Omega^{2}G_{MN}\left[  -8\delta^{\prime
}\left(  W\right)  -4W\delta^{\prime\prime}\left(  W\right)  \right]  =0,$
that follows from the identity $w\delta^{\prime\prime}\left(  w\right)
=-2\delta^{\prime}\left(  w\right)  $, has been added to $V_{MN}$ to obtain
the forms of $V_{MN}^{\left(  1\right)  },V_{MN}^{\left(  2\right)  }$ as shown.

Next, taking into account the remarks in the footnote\footnote{An expression
of the form $A\left(  w\right)  \delta\left(  w\right)  +B\left(  w\right)
\delta^{\prime}\left(  w\right)  +C\left(  w\right)  \delta^{\prime\prime
}\left(  w\right)  =0,$ as in (\ref{Vtot}), is equivalent to three equations
since $\delta\left(  w\right)  ,\delta^{\prime}\left(  w\right)
,\delta^{\prime\prime}\left(  w\right)  $ are three separate distributions. To
carefully separate the equations one considers the Taylor expansion in powers
of $w,$ such as $C\left(  w\right)  =C\left(  0\right)  +C^{\prime}\left(
0\right)  w+\frac{1}{2}C^{\prime\prime}\left(  0\right)  w^{2}+\cdots,$ and
similarly for $B\left(  w\right)  $ and $A\left(  w\right)  $. Then by using
the properties of the delta function as a distribution (i.e. under integration
with smooth functions) $w\delta^{\prime}\left(  w\right)  =-\delta\left(
w\right)  $ and $w\delta^{\prime\prime}\left(  w\right)  =-2\delta^{\prime
}\left(  w\right)  $ and $w^{2}\delta^{\prime\prime}\left(  w\right)
=2\delta\left(  w\right)  ,$ we obtain the following three equations:
$C\left(  0\right)  =0,\;B\left(  0\right)  -2C^{\prime}\left(  0\right)
=0,\;$and $A\left(  0\right)  -B^{\prime}\left(  0\right)  +C^{\prime\prime
}\left(  0\right)  =0.$ \label{subtle}}, we refine the three equations of
motion implied by Eq.(\ref{Vtot}). Each field is expanded in powers of
$W\left(  X\right)  .$ For this, imagine parametrizing $X^{M}$ in terms of
some convenient set of coordinates such that $w\equiv W\left(  X\right)  $ is
one of the independent coordinates. Denoting the remaining $d+1$ coordinates
collectively as $u,$ schematically we can write $G_{MN}\left(  X\right)
=G_{MN}\left(  u,w\right)  ,$ $\Omega\left(  X\right)  =\Omega\left(
u,w\right)  $ and $W\left(  X\right)  =w.$ Then we may expand%
\begin{equation}
G_{MN}\left(  u,w\right)  =G_{MN}\left(  u,0\right)  +wG_{MN}^{\prime}\left(
u,0\right)  +\frac{1}{2}w^{2}G_{MN}^{\prime\prime}\left(  u,0\right)
+\cdots\label{expand}%
\end{equation}
and similarly for $\Omega\left(  u,w\right)  =\Omega\left(  u,0\right)
+\cdots.$ In 2T-field theory in flat space, the zeroth order terms analogous
to $G_{MN}\left(  u,0\right)  $ and $\Omega\left(  u,0\right)  $ were the
physical part of the field, while the rest, which we called the
\textquotedblleft remainder\textquotedblright, was gauge freedom, and could be
set to zero. In this paper we will assume that there is a similar
justification for setting the \textit{remainders} to zero (or some other
convenient gauge choice) \textit{after} the variation of the action has been
performed as in (\ref{Vtot}-\ref{Vtot3}). A procedure for dealing with the
remainders in this fashion could be justified in the case of 2T field theory
in flat space\footnote{This was justified in \cite{2tstandardM} by the fact
that there is a more symmetric starting point for 2T field theory in the form
of a BRST gauge field theory \cite{2tbrst2006} analogous to string field
theory. It is after gauge fixing and simplifying the BRST field theory that
one obtains the simpler and more intuitive form of 2T-field theory used in
\cite{2tstandardM}. Then the working procedure for the simpler form was to
first allow all the remainders as part of the simplified action, and only
\textit{after varying the action} set the remainders to zero (or non-zero but
homogeneous). This is the correct procedure in any gauge theory, i.e. do not
forget the variation with respect to the gauge degrees of freedom. It agreed
with the consequences of the original fully gauge invariant BRST gauge field
theory, as well as the covariantly first quantized worldline theory, at the
level of the classical field equations of motion. Possible consequences of the
remainders, if any, at the second quantization level (path integral) were not
fully clarified and this is part of ongoing research. We don't know yet if the
remainder could play a physically relevant role. \label{after}}. In any case,
setting all the remainders to zero is a legitimate solution of the classical
equations of interest in this paper. Proceeding under this assumption, we keep
only the zeroth order terms in the expansions (\ref{expand}). Then, in view of
footnote (\ref{subtle}), the three classical equations of motion implied by
Eq.(\ref{Vtot}) are%
\begin{equation}
\left[  V_{MN}^{\left(  0\right)  }\left(  X\right)  \right]  _{W\left(
X\right)  =0}=0,\;\left[  V_{MN}^{\left(  1\right)  }\left(  X\right)
\right]  _{W\left(  X\right)  =0}=0,\;\left[  V_{MN}^{\left(  2\right)
}\left(  X\right)  \right]  _{W\left(  X\right)  =0}=0.
\end{equation}
We see immediately from Eq.(\ref{Vtot3}) that the equation of motion
$V_{MN}^{\left(  2\right)  }\left(  u,0\right)  =0$%
\begin{equation}
\partial W\cdot\partial W=4W \label{Whom}%
\end{equation}
reproduces the second Sp$\left(  2,R\right)  $ kinematic constraint
(\ref{sp2}), noting that we have already incorporated the first Sp$\left(
2,R\right)  $ kinematic constraint (\ref{sp1}) in the form $V_{M}=\frac{1}%
{2}\partial_{M}W$ as stated in the beginning of section (\ref{GravAction}). We
now turn to the equation of motion Eq.(\ref{Vtot2}) $V_{MN}^{\left(  1\right)
}\left(  u,0\right)  =0$%
\begin{equation}
\left[  G_{MN}\left(  -6+\nabla^{2}W+\partial W\cdot\partial\ln\Omega
^{2}\right)  -\nabla_{M}\partial_{N}W\right]  _{W\left(  X\right)  =0}=0.
\label{GddW}%
\end{equation}
If we can show that $\left(  -6+\nabla^{2}W+\partial W\cdot\partial\ln
\Omega^{2}\right)  =2,$ then (\ref{GddW}) reproduces the third Sp$\left(
2,R\right)  $ constraint (\ref{sp3}-\ref{sp5}). This is proven as follows. The
variation of the action with respect to $\Omega$ produces on-shell conditions
for $\Omega$; among these Eq.(\ref{F1}), $F^{\left(  1\right)  }=0,$ is solved
by $\partial W\cdot\partial\ln\Omega^{2}=8a\left(  6-\nabla^{2}W\right)  .$ We
insert this in (\ref{GddW}) and then contract Eq.(\ref{GddW}) with $G^{MN}$ to
obtain an equation for only $\nabla^{2}W,$ whose solution is a constant
$\nabla^{2}W=6\left(  d+2\right)  \left(  8a-1\right)  \left[  \left(
8a-1\right)  \left(  d+2\right)  +1\right]  ^{-1}.$ Therefore $\partial
W\cdot\partial\ln\Omega^{2}=48a\left[  \left(  8a-1\right)  \left(
d+2\right)  +1\right]  ^{-1}\allowbreak$ is also a constant. $\allowbreak
$These lead to the on-shell value $\left(  -6+\nabla^{2}W+\partial
W\cdot\partial\ln\Omega^{2}\right)  =6\left(  8a-1\right)  \left[  \left(
8a-1\right)  \left(  d+2\right)  +1\right]  ^{-1}\allowbreak$, which takes the
desired value of $\allowbreak2$ provided $a=\frac{d-2}{8\left(  d-1\right)  }$
as given by Eq.(\ref{a}). With this unique $a$ we obtain the on-shell values
\begin{equation}
\left[  \partial W\cdot\partial\ln\Omega^{2}\right]  _{W\left(  X\right)
=0}=-2\left(  d-2\right)  \;,\;\nabla^{2}W=\allowbreak2\left(  d+2\right)
,\;\;\left[  G_{MN}=\frac{1}{2}\nabla_{M}\partial_{N}W\right]  _{W\left(
X\right)  =0}. \label{Gshell}%
\end{equation}
which is precisely the third Sp$\left(  2,R\right)  $ kinematic constraint
(\ref{sp3}-\ref{sp5}).

Hence, we have constructed an action consistent with the Sp$\left(
2,R\right)  $ conditions (\ref{sp1}-\ref{sp5}), and the condition
$Q_{11}=W\left(  X\right)  =0$. These were the necessary kinematic constraints
to remove all the ghosts in the two-time theory for Gravity. They produce a
shadow that describes gravity in $\left(  d-1\right)  +1$ dimensions as in
Eq.(\ref{worldlineGravity}) in the worldline formalism, and also in the field
theory formalism as discussed before \cite{2tfield} and which will be further
explained below.

The remaining field equation $V_{MN}^{\left(  0\right)  }\left(  u,0\right)
=0$ in Eq.(\ref{Vtot1}) now gives the desired dynamical equation that has the
form of Einstein's equation in $d+2$ dimensions
\begin{equation}
\left[  R_{MN}\left(  G\right)  -\frac{1}{2}G_{MN}R\left(  G\right)  \right]
_{W\left(  X\right)  =0}=\left[  T_{MN}\left(  \Omega,G\right)  \right]
_{W\left(  X\right)  =0}, \label{einstein}%
\end{equation}
with an energy-momentum source $T_{MN}\left(  \Omega,G\right)  $ provided by
the dilaton field%
\begin{equation}
T_{MN}=\left[
\begin{array}
[c]{c}%
-\frac{1}{2a}\left(  \partial_{M}\ln\Omega\right)  \left(  \partial_{N}%
\ln\Omega\right)  +\frac{1}{2}G_{MN}\left(  \frac{1}{2a}\partial\ln\Omega
\cdot\partial\ln\Omega-\frac{V\left(  \Omega\right)  }{\Omega^{2}}\right) \\
-\frac{1}{\Omega^{2}}\left(  G_{MN}\nabla^{2}\Omega^{2}-\nabla_{M}\partial
_{N}\Omega^{2}\right)
\end{array}
\right]
\end{equation}
The unique value of the constant $a$ (\ref{a}) will be required also by
additional Sp$\left(  2,R\right)  $ relations as will be seen below. Under the
assumption that the dilaton field $\Omega$ is invertible (certainly so if it
has a nonzero vacuum expectation value), we have divided by the field $\Omega$
to extract $T_{MN}.$ Once all the kinematic constraints obtained above and
below are taken into account, this correctly reduces to General Relativity in
$d$ dimensions as a shadow (see below). So, $S=S_{G}+S_{\Omega}+S_{W}$ is a
consistent action that produces the correct gravitational classical field
equations directly in $d+2$ dimensions.

\section{Equations of motion for $\Omega$ \label{eomOmega}}

We now turn to the variation of the action with respect to the dilaton
$\Omega$ to extract its equations of motion. After integration by parts that
produce $\delta^{\prime}\left(  W\right)  ,\delta^{\prime\prime}\left(
W\right)  $ terms, we obtain
\begin{align}
\delta_{\Omega}\left(  S_{\Omega}\right)   &  =\gamma\int d^{d+2}X~\sqrt
{G}\delta\Omega~\left\{  \delta\left(  W\right)  \left(  -\frac{1}{a}%
\nabla^{2}\Omega-V^{\prime}\left(  \Omega\right)  \right)  -\frac{1}{a}%
\delta^{\prime}\left(  W\right)  \partial W\cdot\partial\Omega\right\}  ,\\
\delta_{\Omega}\left(  S_{W}\right)   &  =\gamma\int d^{d+2}X~~\sqrt{G}%
\delta\Omega^{2}\left\{  \delta^{\prime}\left(  W\right)  \left(  4-\nabla
^{2}W\right)  -\nabla\cdot\left(  \partial W\delta^{\prime}\left(  W\right)
\right)  \right\} \\
&  =\gamma\int d^{d+2}X~\sqrt{G}\delta\Omega~\left\{
\begin{array}
[c]{c}%
\delta^{\prime}\left(  W\right)  \Omega\left(  24-4\nabla^{2}W\right) \\
+\delta^{\prime\prime}\left(  W\right)  \Omega\left(  -2\partial
W\cdot\partial W+8W\right)
\end{array}
\right\}  ,
\end{align}
where we have added the vanishing expression $\Omega\left[  16\delta^{\prime
}\left(  W\right)  +8W\delta^{\prime\prime}\left(  W\right)  \right]  =0$ to
obtain a convenient form. Including $\delta_{\Omega}\left(  S_{G}\right)  ,$
which contains only $\delta\left(  W\right)  ,$ we obtain the total variation
$\delta_{\Omega}\left(  S_{\Omega}+S_{W}+S_{G}\right)  =\gamma\int
d^{d+2}X~\sqrt{G}\delta\Omega F\left(  X\right)  ,$ which gives the equation
of motion $F=0$%
\begin{align}
F  &  \equiv\delta\left(  W\right)  F^{\left(  0\right)  }+\delta^{\prime
}\left(  W\right)  F^{\left(  1\right)  }+\delta^{\prime\prime}\left(
W\right)  F^{\left(  2\right)  }=0\label{Ftot}\\
F^{\left(  0\right)  }  &  \equiv2R\Omega-\frac{1}{a}\nabla^{2}\Omega
-V^{\prime}\left(  \Omega\right) \label{F0}\\
F^{\left(  1\right)  }  &  \equiv-\frac{1}{a}\partial W\cdot\partial
\Omega+4\Omega\left(  6-\nabla^{2}W\right) \label{F1}\\
F^{\left(  2\right)  }  &  \equiv-2\Omega\left[  \partial W\cdot\partial
W-4W\right]  \label{F2}%
\end{align}
As in the discussion before, we seek a solution when the remainders of the
fields vanish. Then the three on-shell equations are $F^{\left(  0\right)
}=F^{\left(  1\right)  }=F^{\left(  2\right)  }=0.$ The expression $F^{\left(
2\right)  }=0$ is satisfied since it is identical to Eq.(\ref{Whom}) which
amounts to the Sp$\left(  2,R\right)  $ kinematic constraints (\ref{sp1}%
-\ref{sp2}). The condition $F^{\left(  1\right)  }=0$ produces a kinematic
constraint $\partial W\cdot\partial\ln\Omega^{2}=8a\left(  6-\nabla
^{2}W\right)  $ for the field $\Omega$ as used in the derivation of
Eq.(\ref{Gshell}). After inserting the on-shell value $\nabla^{2}W=2\left(
d+2\right)  $ from Eq.(\ref{Gshell}) for the spacial value of $a,$ the
constraint becomes%
\begin{equation}
F^{\left(  1\right)  }=\left[  \partial W\cdot\partial\Omega+\left(
d-2\right)  \Omega\right]  _{W\left(  X\right)  =0}=0. \label{kinPhi}%
\end{equation}
In the flat limit of Eq.(\ref{flat}) this reduces to $F_{flat}^{\left(
1\right)  }=\left[  2X\cdot\partial+\left(  d-2\right)  \right]  \Omega=0,$
which is a homogeneity constraint on $\Omega$ consistent with the assigned
dimension of the field $\Omega$ in Eq.(\ref{dims}). Therefore, this is another
consistency condition that requires the value of $a$ in Eq.(\ref{a}). We will
see below when we study variations with respect to the field $W,$ that there
is a stronger and independent gauge symmetry argument that fixes uniquely the
same value of $a.$

The dynamical equation for $\Omega$ is now determined by setting $F^{\left(
0\right)  }=0$ with the special $a$
\begin{equation}
\left[  \nabla^{2}\Omega+\frac{d-2}{8\left(  d-1\right)  }\left(  V^{\prime
}\left(  \Omega\right)  -2\Omega R\left(  G\right)  \right)  \right]
_{W\left(  X\right)  =0}=0. \label{confKG}%
\end{equation}
Here there is an interesting point to be emphasized. The precise coefficient
of $\Omega R$ (which is $2a$) is the one that would normally appear for the
conformal scalar in $d$ dimensions, but note that the Laplacian and the
curvature $R\left(  G\right)  $ in our case are in $d+2$ dimensions not in $d$
dimensions. If the coefficient had been the one appropriate for $d+2$
dimensions, namely $-\frac{d}{4\left(  d+1\right)  },$ then there would have
been a local Weyl symmetry that could eliminate $\Omega\left(  X\right)  $
from the theory by a local Weyl rescaling. However, this is not the case
presently. Nevertheless, we will identify later an accidental local Weyl
symmetry for the \textquotedblleft conformal shadow\textquotedblright\ in $d$
dimensions (that is, not Weyl in the full $d+2$ dimensions). This partially
local \textquotedblleft accidental\textquotedblright\ Weyl symmetry\ will
indeed eliminate the fluctuations of $\Omega\left(  X\right)  $ in the shadow
subspace, but still keeping some dependence of $\Omega$ in the extra
dimensions. In this way, the special value of $a$ will allow us to eliminate
the massless Goldstone boson that arises due to spontaneous breakdown of scale
invariance in the shadow subspace.

\section{Equations of motion for $W$ \label{eomW}}

The part of the action $S_{G}+S_{\Omega}$ contains $W$ only in the delta
function, so its variation is proportional to $\delta^{\prime}\left(
W\right)  $%
\begin{equation}
\delta_{W}\left(  S_{G}+S_{\Omega}\right)  =\gamma\int d^{d+2}X\sqrt{G}\left(
\delta W\right)  \delta^{\prime}\left(  W\right)  \left[  \Omega^{2}R\left(
G\right)  +\frac{1}{2a}\partial\Omega\cdot\partial\Omega-V\left(
\Omega\right)  \right]
\end{equation}
Varying $W$ in $S_{W}$ produces terms proportional to $\delta^{\prime}\left(
W\right)  ,\delta^{\prime\prime}\left(  W\right)  $ and $\delta^{\prime
\prime\prime}\left(  W\right)  $ as follows%
\begin{align}
\delta_{W}\left(  S_{W}\right)   &  =\gamma\int d^{d+2}X\sqrt{G}~\delta
W\left\{
\begin{array}
[c]{c}%
\delta^{\prime\prime}\left(  W\right)  \left[  \Omega^{2}\left(  4-\nabla
^{2}W\right)  +\partial W\cdot\partial\Omega^{2}\right] \\
-\nabla\cdot\partial\left[  \Omega^{2}\delta^{\prime}\left(  W\right)
\right]  -\nabla\cdot\left[  \delta^{\prime}\left(  W\right)  \partial
\Omega^{2}\right]
\end{array}
\right\} \\
&  =\gamma\int d^{d+2}X\sqrt{G}~\delta W\left\{
\begin{array}
[c]{l}%
\delta^{\prime}\left(  W\right)  \left[  -2\nabla^{2}\Omega^{2}\right] \\
+\delta^{\prime\prime}\left(  W\right)  \left[  \Omega^{2}\left(
16-2\nabla^{2}W\right)  -2\partial W\cdot\partial\Omega^{2}\right] \\
+\delta^{\prime\prime\prime}\left(  W\right)  \Omega^{2}\left[  -\partial
W\cdot\partial W+4W\right]
\end{array}
\right\}
\end{align}
We have added the vanishing expression $\Omega^{2}\left[  12\delta
^{\prime\prime}\left(  W\right)  +4W\delta^{\prime\prime\prime}\left(
W\right)  \right]  =0$ to obtain a convenient form. Thus the $\delta_{W}$
variation of the total action has the form $\delta_{W}\left(  S_{G}+S_{\Omega
}+S_{W}\right)  =\gamma\int d^{d+2}X\sqrt{G}~\delta W~Z\left(  X\right)  ,$
which leads to the equation of motion $Z\left(  X\right)  =0$%
\begin{align}
Z  &  \equiv\delta^{\prime}\left(  W\right)  Z^{\left(  1\right)  }%
+\delta^{\prime\prime}\left(  W\right)  Z^{\left(  2\right)  }+\delta
^{\prime\prime\prime}\left(  W\right)  Z^{\left(  3\right)  }=0,\label{Otot}\\
Z^{\left(  1\right)  }  &  \equiv\Omega^{2}R\left(  G\right)  -2\nabla
^{2}\Omega^{2}+\frac{1}{2a}\partial\Omega\cdot\partial\Omega-V\left(
\Omega\right)  ,\label{O1}\\
Z^{\left(  2\right)  }  &  \equiv\Omega^{2}\left(  16-2\nabla^{2}W\right)
-2\partial W\cdot\partial\Omega^{2},\label{O2}\\
Z^{\left(  3\right)  }  &  \equiv-\Omega^{2}\left[  \partial W\cdot\partial
W-4W\right]  . \label{O3}%
\end{align}
It is remarkable that, if we use the on-shell \textit{kinematic} equations of
motion for $W$ and $\Omega$ (\ref{Whom},\ref{Gshell},\ref{kinPhi}) we get
$\left[  Z^{\left(  2\right)  }\right]  _{W=0}=Z^{\left(  3\right)  }=0$.
Then, if we also use the \textit{dynamical} equations for both $G_{MN}$ and
$\Omega$ (\ref{einstein},\ref{confKG}), we also obtain $\left[  Z^{\left(
1\right)  }\right]  _{W=0}=0.$ These remarkable identities are possible
\textit{only if }$a$\textit{ has precisely the special value in Eq.(\ref{a})}.

Therefore minimizing the action with respect to $W$ does not produce any new
kinematic or dynamical on-shell conditions for the fields. Hence, the on-shell
value of $W\left(  X\right)  $ is arbitrary, indicating the presence of a
gauge symmetry only for the special value of $a=\frac{d-2}{8\left(
d-1\right)  }.$

\section{Off-shell gauge symmetry \label{local}}

Let us now prove that indeed there is an off-shell gauge symmetry without
using any of the kinematic or the dynamical equations of motion. A gauge
transformation of the total action has the form $\delta_{\Lambda}S=\gamma\int
d^{d+2}X\sqrt{G}\left(  V_{MN}\delta_{\Lambda}G^{MN}+F\delta_{\Lambda}%
\Omega+Z\delta_{\Lambda}W\right)  $ where $V_{MN},F,Z$ are given in
Eqs.(\ref{Vtot},\ref{Ftot},\ref{Otot}) respectively, but taken off-shell. We
explore a gauge transformation of the form%
\begin{equation}
\delta_{\Lambda}G^{MN}=\alpha G^{MN},\;\delta_{\Lambda}\Omega=\beta
\Omega,\;\delta_{\Lambda}W=\Lambda W. \label{gaugeSymm}%
\end{equation}
with local functions $\alpha\left(  X\right)  ,\beta\left(  X\right)  $ that
will be determined below in terms of $\Lambda\left(  X\right)  $. We collect
the coefficients of $\delta\left(  W\right)  ,\delta^{\prime}\left(  W\right)
,\delta^{\prime\prime}\left(  W\right)  $ in the gauge transformation
$\delta_{\Lambda}S$ after using the delta function identities $w\delta
^{\prime}\left(  w\right)  =-\delta\left(  w\right)  ,~w\delta^{\prime\prime
}\left(  w\right)  =-2\delta^{\prime}\left(  w\right)  $ and $w\delta
^{\prime\prime\prime}\left(  w\right)  =-3\delta^{\prime\prime}\left(
w\right)  .$ This gives
\begin{equation}
V_{MN}\delta_{\Lambda}G^{MN}+F\delta_{\Lambda}\Omega+Z\delta_{\Lambda
}W=\left\{
\begin{array}
[c]{c}%
\delta\left(  W\right)  \left[  \alpha G^{MN}V_{MN}^{\left(  0\right)  }%
+\beta\Omega F^{\left(  0\right)  }-\Lambda Z^{\left(  1\right)  }\right] \\
+\delta^{\prime}\left(  W\right)  \left[  \alpha G^{MN}V_{MN}^{\left(
1\right)  }+\beta\Omega F^{\left(  1\right)  }-2\Lambda Z^{\left(  2\right)
}\right] \\
+\delta^{\prime\prime}\left(  W\right)  \left[  \alpha G^{MN}V_{MN}^{\left(
2\right)  }+\beta\Omega F^{\left(  2\right)  }-3\Lambda Z^{\left(  3\right)
}\right]
\end{array}
\right\}  \label{delGamma}%
\end{equation}
We first analyze the term proportional to $\delta^{\prime\prime}\left(
W\right)  .$ After inserting the off-shell quantities $V_{MN}^{\left(
2\right)  },F^{\left(  2\right)  },Z^{\left(  3\right)  }$ if Eqs.(\ref{Vtot3}%
,\ref{F2},\ref{O3}) we see that the $\delta^{\prime\prime}\left(  W\right)  $
term can be written as a total divergence\footnote{Use the identity
$\nabla\cdot\left[  \partial W\delta^{\prime}\left(  W\right)  A\Phi
^{2}\right]  =\delta^{\prime\prime}\left(  W\right)  \left(  \partial
W\cdot\partial W-4W\right)  A\Phi^{2}+\delta^{\prime}\left(  W\right)  \left[
\nabla\cdot\left(  \partial WA\Phi^{2}\right)  -8A\Phi^{2}\right]  .$} plus a
term proportional to $\delta^{\prime}\left(  W\right)  :$
\begin{align}
&  \delta^{\prime\prime}\left(  W\right)  \left[  \alpha G^{MN}V_{MN}^{\left(
2\right)  }+\beta\Omega F^{\left(  2\right)  }-3\Lambda Z^{\left(  3\right)
}\right] \nonumber\\
&  =\delta^{\prime\prime}\left(  W\right)  \Omega^{2}\left(  \partial
W\cdot\partial W-4W\right)  \left(  \frac{\alpha}{2}\left(  d+2\right)
-2\beta+3\Lambda\right) \\
&  =\nabla\cdot\left[  \partial W\delta^{\prime}\left(  W\right)  \left(
\frac{\alpha}{2}\left(  d+2\right)  -2\beta+3\Lambda\right)  \Omega
^{2}\right]  +U^{\left(  1\right)  }\delta^{\prime}\left(  W\right)
\end{align}
where%
\[
U^{\left(  1\right)  }\left(  \alpha,\beta,\Lambda\right)  =\left\{
\begin{array}
[c]{c}%
\Omega^{2}\left(  \frac{\alpha}{2}\left(  d+2\right)  -2\beta+3\Lambda\right)
\left(  8-\nabla^{2}W\right) \\
-\partial W\cdot\partial\left[  \Omega^{2}\left(  \frac{\alpha}{2}\left(
d+2\right)  -2\beta+3\Lambda\right)  \right]
\end{array}
\right\}  .
\]
The total divergence can be dropped in $\delta_{\Lambda}S$ since $\int
d^{d+2}X\sqrt{G}\left(  \nabla\cdot Q\right)  =\int d^{d+2}X\partial
_{M}\left(  \sqrt{G}G^{MN}Q_{N}\right)  \rightarrow0.$ Therefore, in the gauge
transformation (\ref{delGamma}) the part proportional $\delta^{\prime\prime
}\left(  W\right)  $ can be eliminated at the expense of adding $U^{\left(
1\right)  }\delta^{\prime}\left(  W\right)  $ to the part proportional to
$\delta^{\prime}\left(  W\right)  .$ Now we have 3 functions $\left(
\alpha,\beta,\Lambda\right)  $ at our disposal to fix to zero the 2 remaining
terms of the gauge transformation (\ref{delGamma}), namely%
\begin{align}
0  &  =\alpha G^{MN}V_{MN}^{\left(  0\right)  }+\beta\Omega F^{\left(
0\right)  }-\Lambda Z^{\left(  1\right)  },\\
0  &  =\alpha G^{MN}V_{MN}^{\left(  1\right)  }+\beta\Omega F^{\left(
1\right)  }-2\Lambda Z^{\left(  2\right)  }+U^{\left(  1\right)  }\left(
\alpha,\beta,\Lambda\right)  .
\end{align}
Clearly there is freedom to fix $\alpha,\beta$ in terms of an arbitrary
$\Lambda$ to insure the \textit{off-shell gauge symmetry} of the action
$\delta_{\Lambda}S=0.$

The analysis of the equations of motion in the previous section had indicated
that $W\left(  X\right)  $ was arbitrary on-shell. The discussion in this
section shows that this freedom extends to also off-shell, since according to
(\ref{gaugeSymm}), we can use the gauge freedom $\Lambda\left(  X\right)  $ to
choose $W\left(  X\right)  $ arbitrarily as a function of $X.$

\section{General Relativity as a shadow}

From the gauge transformations (\ref{gaugeSymm}) we see that the gauge
symmetry indicates that $W\left(  X\right)  $ is gauge freedom, so it can be
chosen arbitrarily as a function of $X^{M}$ before restricting spacetime by
the condition $W\left(  X\right)  =0$ in $d+2$ dimensions. This freedom is
related to the production of multiple $d$ dimensional shadows of the same
$d+2$ dimensional system.

Our action is also manifestly invariant under general coordinate
transformations in $d+2$ dimensions, which can be used to fix components of
the metric $G_{MN}\left(  X\right)  .$ This freedom will also be used in the
production of shadows.

To proceed to generate a shadow of our theory in $d$ dimensions it is useful
to choose a parametrization of the coordinates $X^{M}$ in $d+2$ dimensions in
such a way as to embed a $d$ dimensional subspace $x^{\mu}$ in the higher
space $X^{M}$. There are many ways of doing this, to create various shadows
with different meanings of \textquotedblleft time\textquotedblright\ as
perceived by observers that live in the fixed shadow $x^{\mu}$. This was
discussed in the past for the particle level of 2T-physics and recently for
the field theory level \cite{emergentfieldth1}\cite{emergentfieldth2}. A
particular parametrization which is useful to explain massless particles and
conformal symmetry in flat space \cite{Dirac}-\cite{salam} as a shadow of
Lorentz symmetry in flat $\left(  d+2\right)  $ dimensions was commonly used
in our past work. We will call this the \textquotedblleft conformal
shadow\textquotedblright. The parametrization in this section, which should be
understood to correspond to one particular shadow, is a generalization of the
conformal shadow to curved space.

We choose a parametrization of $X^{M}$ in terms of $d+2$ coordinates named
$\left(  w,u,x^{\mu}\right)  $. In the new curved space $\left(  w,u,x^{\mu
}\right)  ,$ where the basis is specified by $\partial_{M}=\left(
\partial_{w},\partial_{u},\partial_{\mu}\right)  ,$ we use general coordinate
transformations to gauge fix $d+2$ functions among the $G^{MN}\left(
w,u,x^{\mu}\right)  ,$ namely $G^{wu}=1,$ $G^{uu}=G^{w\mu}=0,$ so that the
metric takes the following form%
\begin{equation}
G^{MN}=%
\begin{array}
[c]{cc}%
M\backslash N &
\begin{array}
[c]{ccc}%
~w~ & ~u~ & ~~~\nu
\end{array}
\;\;\;\\%
\begin{array}
[c]{c}%
w\\
u\\
\mu
\end{array}
& \left(
\begin{array}
[c]{ccc}%
G^{ww} & -1 & 0\\
-1 & 0 & G^{u\nu}\\
0 & G^{\mu u} & G^{\mu\nu}%
\end{array}
\right)
\end{array}
\label{GMN}%
\end{equation}
In this basis we make a choice for $W\left(  X\right)  $ which specifies the
conformal shadow. Namely we take $W\left(  X\right)  =w$ as one of the
coordinates
\begin{equation}
W\left(  X\right)  =W\left(  w,u,x^{\mu}\right)  =w \label{Wchoice}%
\end{equation}
We compute $\partial_{M}W\left(  X\right)  $ in this basis and find
\begin{equation}
\partial_{M}W=\left(  1,0,0\right)  _{M}.
\end{equation}
Now we apply the Sp$\left(  2,R\right)  $ \textit{kinematical} constraint
$4W=\partial W\cdot\partial W,$ derived from field theory in Eq.(\ref{Whom})
or from the worldline theory in (\ref{sp1},\ref{sp2})%
\begin{equation}
4W=G^{MN}\partial_{M}W\partial_{N}W=G^{MN}\left(  1,0,0\right)  _{M}\left(
1,0,0\right)  _{N}=G^{ww}.
\end{equation}
This determines
\begin{equation}
G^{ww}\left(  w,u,x^{\mu}\right)  =4w. \label{Gww}%
\end{equation}
Next we apply the Sp$\left(  2,R\right)  $ \textit{kinematical} constraint
(\ref{sp5}) which was also derived in field theory in Eq.(\ref{Gshell}). We
will use the equivalent form in (\ref{sp4}), $-2G^{MN}=V^{K}\partial_{K}%
G^{MN}-\partial_{K}V^{M}G^{KN}-\partial_{K}V^{N}G^{MK},$ where we insert
$V^{M}$ as obtained from (\ref{sp1})
\begin{equation}
V^{M}\left(  w,u,x^{\mu}\right)  =\frac{1}{2}G^{MN}\partial_{N}W=\left(
2w,-\frac{1}{2},0\right)  ^{M}. \label{VM}%
\end{equation}
Then we get $V^{M}\partial_{M}=\left(  2w\partial_{w}-\frac{1}{2}\partial
_{u}\right)  ,$ and the kinematic constraint (\ref{sp4}) takes the form%
\begin{equation}
-2G^{MN}=\left(  2w\partial_{w}-\frac{1}{2}\partial_{u}\right)  G^{MN}%
-2\delta_{w}^{M}G^{wN}-2\delta_{w}^{N}G^{Mw}.
\end{equation}
We check that $G^{ww}=4w,$ $G^{wu}=1,~G^{uu}=G^{w\mu}=0,$ all satisfy these
kinematical conditions automatically, while the remaining components, $G^{\mu
u},G^{\mu\nu},$ must depend on $u,x$ and $w$ only in the following specific
form%
\begin{align}
G^{\mu\nu}\left(  w,u,x^{\mu}\right)   &  =e^{4u}\hat{g}^{\mu\nu}\left(
x,e^{4u}w\right)  ,\;\;\\
G^{\mu u}\left(  w,u,x^{\mu}\right)   &  =e^{4u}\gamma^{\mu}\left(
x,e^{4u}w\right)  .
\end{align}

As explained following Eq.(\ref{expand}), in an expansion in powers
of $w$ only the zeroth order term is kept in our solution. So, for
our purposes here $G^{\mu\nu}\left(  w,u,x^{\mu}\right)
=e^{4u}g^{\mu\nu}\left(  x\right)  $ and $G^{\mu u}\left(
w,u,x^{\mu}\right)  =e^{4u}\gamma^{\mu}\left(  x\right) $ are
independent of $w.$ Even though we have already used up all of the
gauge freedom of general coordinate transformations to fix $d+2$
functions of $\left(  w,u,x^{\mu}\right)  $ as in Eq.(\ref{GMN}),
there still remains general coordinate symmetry to reparameterize
arbitrarily the subspace $\left( u,x^{\mu}\right)  $ in such a way
that the form of the metric in Eq.(\ref{GMN}) remains unchanged.
This allows us to fix $d$ functions of $\left(  u,x^{\mu}\right)  $
arbitrarily as gauge choices. Therefore, for the
$w$ independent components of the metric at $w=0$ we can make the gauge choice%
\begin{equation}
G^{\mu u}\left(  0,u,x^{\mu}\right)  =0,\;\rightarrow\;\gamma^{\mu}\left(
x\right)  =0. \label{Gmuu}%
\end{equation}
We remain only with the degrees of freedom of the metric $g^{\mu\nu}\left(
x\right)  $ in $d$ dimensions given by%
\begin{equation}
G^{\mu\nu}\left(  0,u,x^{\mu}\right)  =e^{4u}g^{\mu\nu}\left(  x\right)  .
\label{Gmunu}%
\end{equation}
There still remains gauge symmetry for general coordinate transformations in
the $x^{\mu}$ subspace. In this form it is easy to compute the determinant of
$G^{MN},$ given in (\ref{GMN}). This gives $\det\left(  G^{-1}\right)
=-e^{4du}\det\left(  g^{-1}\left(  x\right)  \right)  ,$ or
\begin{equation}
\sqrt{G\left(  w,u,x^{\mu}\right)  }=e^{-2du}\sqrt{-g\left(  x\right)  }.
\label{sqrG}%
\end{equation}
As a final check we compute that $\nabla^{2}W=2\left(  d+2\right)  $ is also
satisfied as required by Eq.(\ref{Gshell}), as follows%
\begin{align}
\nabla^{2}W  &  =\frac{1}{\sqrt{G}}\partial_{M}\left(  \sqrt{G}G^{MN}%
\partial_{N}W\right)  =\frac{1}{\sqrt{G}}\partial_{M}\left(  \sqrt{G}%
G^{Mw}\partial_{w}w\right) \\
&  =\frac{1}{\sqrt{G}}\partial_{w}\left(  \sqrt{G}G^{ww}\right)  +\frac
{1}{\sqrt{G}}\partial_{u}\left(  \sqrt{G}G^{uw}\right) \\
&  =\partial_{w}\left(  4w\right)  -e^{2du}\partial_{u}e^{-2du}=4+2d.
\end{align}
The metric $G^{MN}\left(  X\right)  $ given in Eqs.(\ref{GMN},\ref{Gww}%
,\ref{Gmuu},\ref{Gmunu}) shows that, after imposing the kinematic constraints
at the classical level, the conformal shadow is described only in terms of the
degrees of freedom $g_{\mu\nu}\left(  x\right)  $ in $d$ dimensions.

We now go through similar arguments to impose the kinematic constraint
(\ref{kinPhi}) for $\Omega.$ This takes the form
\begin{equation}
0=\left(  V^{M}\partial_{M}+\frac{d-2}{2}\right)  \Omega=\left(
2w\partial_{w}-\frac{1}{2}\partial_{u}+\frac{d-2}{2}\right)  \Omega\left(
w,u,x\right)
\end{equation}
The solution is, $\Omega\left(  w,u,x\right)  =e^{-\left(  d-2\right)  u}%
\hat{\phi}\left(  x,e^{-4u}w\right)  ,$ in which the zeroth order term in the
expansion in powers of $w$ is identified as the physical field $\phi\left(
x\right)  $ in $d$ dimensions
\begin{equation}
\Omega\left(  0,u,x\right)  =e^{\left(  d-2\right)  u}\phi\left(  x\right)  .
\label{PhiSolved}%
\end{equation}

After solving the kinematic constraints we have arrived at the conformal
shadow with only the degrees of freedom $g^{\mu\nu}\left(  x\right)
,\phi\left(  x\right)  .$ We can now evaluate the full action for the shadow.
The volume element becomes%
\begin{equation}
d^{d+2}X\sqrt{G}\delta\left(  W\left(  X\right)  \right)  =dwdu\left(
d^{d}x\right)  \sqrt{-g\left(  x\right)  }e^{-2du}\delta\left(  w\right)
\end{equation}
Every term in the Lagrangian density is now independent of $w$ and
has the same overall factor $e^{2du}$ as the only possible
dependence on $u.$ Specifically $\Omega^{2}$ is proportional to
$e^{2\left(  d-2\right)  u}$ and $R\left(  G\right)  $ is
proportional to $e^{4u},$ so $\Omega^{2}R\left( G\right)  $ is
proportional to $e^{2du},$ etc. Both the $w$ and $u$ dependences are
explicit. So the action in $d+2$ dimensions produces the
following shadow action in $d$ dimensions%
\begin{equation}
S_{G}+S_{\Omega}+S_{W}=\left(  \gamma\int du\right)  \int\left(
d^{d}x\right)  \sqrt{-g\left(  x\right)  }L_{d}\left(  x\right)  ,
\label{gamma}%
\end{equation}
where the overall renormalization constant $\gamma$ is chosen so that $\left(
\gamma\int du\right)  =1.$ The factor of $\gamma$ can be interpreted as a
renormalization of Planck's constant $\hbar$ since in the path integral
$\hbar$ appears only in the form $S/\hbar.$

The shadow Lagrangian in $d$ dimensions $L_{d}\left(  x\right)  $ takes the
form
\begin{equation}
S\left(  g,\phi\right)  =\int d^{d}x\sqrt{-g}\left(  \frac{1}{2a}g^{\mu\nu
}\partial_{\mu}\phi\partial_{\nu}\phi+R\phi^{2}-V\left(  \phi\right)  \right)
. \label{confScalar}%
\end{equation}
Recall that the special value of $a$ was \textit{required }to generate
consistently all of the Sp$\left(  2,R\right)  $ kinematic constraints. Then
$\phi\left(  x\right)  $ is the \textit{conformal }scalar in $d$ dimensions.
As discussed earlier following Eq.(\ref{confKG}), this action has an
accidental local Weyl symmetry given by $S\left(  \tilde{g},\tilde{\phi
}\right)  =S\left(  g,\phi\right)  $ under the gauge transformation
\begin{equation}
\tilde{g}_{\mu\nu}\left(  x\right)  =e^{2\lambda\left(  x\right)  }g_{\mu\nu
}\left(  x\right)  ,\;\tilde{\phi}\left(  x\right)  =e^{-\frac{d-2}{2}%
\lambda\left(  x\right)  }\phi\left(  x\right)  .
\end{equation}
This gauge freedom can be used to gauge fix $\phi\left(  x\right)  $ except
for an overall constant that absorbs dimensions. Assuming $\phi\left(
x\right)  $ has a non-zero vacuum expectation value $\phi_{0}$, we may write
$\phi^{2}\left(  x\right)  =\phi_{0}^{2}e^{\left(  d-2\right)  \sigma\left(
x\right)  }$ and gauge fix the fluctuation $\sigma\left(  x\right)  =0$. Note
that $\sigma\left(  x\right)  $ would have been the Goldstone boson for
dilatations, but in the present theory it is not a physical degree of freedom.

We can try to trace back the origin of this accidental Weyl symmetry. It is
related to the gauge symmetry discussed in section (\ref{local}). That
symmetry was already used to gauge fix $W\left(  X\right)  =w.$ There remains
leftover gauge symmetry that does not change $w,$ but can change the $w$
independent parts of the fields $\Omega,G_{MN}$ which describe the shadow. So,
the conformal shadow ends up having the accidental Weyl symmetry.

It is important to emphasize that the action in $d+2$ dimensions does not have
a Weyl symmetry, therefore $\Omega$ could not be removed locally. In fact, as
seen from (\ref{PhiSolved}), even after gauge fixing $\phi\left(  x\right)  ,$
as well putting the theory on shell, the original field becomes $\Omega\left(
w,u,x\right)  =e^{\left(  d-2\right)  u}\hat{\phi}\left(  x,e^{4u}w\right)
=e^{\left(  d-2\right)  u}\phi_{0}+O\left(  w\right)  ,$ so even on-shell it
still depends on the spacetime coordinate $u$ in $d+2$ dimensions (also on $w$
before setting $w=0$). Thus, the full $\Omega$ is not a trivial pure gauge
freedom in our theory.

The shadow that emerged with a constant $\phi_{0}$ has exactly the form of
General Relativity with a possible cosmological constant contributed by
$\phi_{0}^{-2}V\left(  \phi_{0}\right)  ,$ if this quantity is non-vanishing%
\begin{equation}
S\left(  g,\phi_{0}\right)  =\int d^{d}x\sqrt{-g}\left(  \phi_{0}^{2}R\left(
g\right)  -V\left(  \phi_{0}\right)  \right)  . \label{shadowGrav}%
\end{equation}
What is left behind from $\phi\left(  x\right)  $ in the shadow is only the
constant $\phi_{0}$ of mass scale $M^{\frac{d-2}{2}}.$ This constant cannot be
determined within the theory we have outlined so far. With our potential
$V\left(  \phi\right)  $ in Eq.(\ref{Vphi}), minimizing the action with
respect to $\phi\left(  x\right)  ,$ and then gauge fixing to $\phi\left(
x\right)  =\phi_{0}$, does not produce a new equation for $\phi_{0}$ other
than the one obtained by minimizing the action with respect to the metric
$g_{\mu\nu},$ namely $R\left(  g\right)  =\frac{1}{2\phi_{0}}V^{\prime}\left(
\phi_{0}\right)  =\lambda\phi_{0}^{4/\left(  d-2\right)  }$. An effective
potential $V\left(  \phi\right)  $ with a non-trivial minimum could determine
$\phi_{0}$. We assume that a non-trivial minimum arises self-consistently from
either quantum fluctuations (dimensional transmutation \cite{colemanweinberg}%
), or from the completion of our theory into string theory or M-theory (with 2
times). Although we could not determine $\phi_{0}\sim M^{\frac{d-2}{2}}$
within the classical considerations here, this $\phi_{0}$ that appears as a
constant shadow of $\Omega\left(  X\right)  $ to observers in $x$-space, is
evidently related to Newton's constant $G_{d}$ or the Planck constant
$\kappa_{d}$ or the Planck scale $l_{p}$ in $d$ dimensions%
\begin{equation}
\phi_{0}^{2}=\frac{1}{16\pi G_{d}}=\frac{1}{2\kappa_{d}^{2}}=\frac{2\pi
}{\left(  2\pi l_{p}\right)  ^{d-2}}\sim M^{d-2}. \label{gravConstant}%
\end{equation}

\section{Gravitational non-constant, new cosmology?}

We now outline the coupling of our gravity triplet $\left(  W,\Omega
,G^{MN}\right)  $ to matter fields of the type Klein-Gordon ($S_{i}\left(
X\right)  $), Dirac ($\Psi\left(  X\right)  $) and Yang-Mills ($A_{M}\left(
X\right)  $). In flat 2T field theory these must have the following
engineering dimensions \cite{2tstandardM}
\begin{equation}
\dim\left(  X^{M}\right)  =1,\;\dim\left(  S_{i}\right)  =-\frac{d-2}%
{2},\;\dim\left(  \Psi\right)  =-\frac{d}{2},\;\dim\left(  A_{M}\right)  =-1
\end{equation}
The general 2T field theory of these fields in flat space in $d+2$ dimensions
was given in \cite{2tstandardM}. The matter part of the theory in curved space
follows from the flat theory in \cite{2tstandardM} by making the substitutions
indicated in Table-1.
\begin{gather*}%
\begin{tabular}
[c]{|l|l|l|}\hline
Quantity & Flat & Curved\\\hline
metric & $\eta^{MN}$ & $G^{MN}\left(  X\right)  $\\
volume element~ & $\left(  d^{d+2}X\right)  \delta\left(  X^{2}\right)  \;\;$
& $\left(  d^{d+2}X\right)  \sqrt{G}\delta\left(  W\left(  X\right)  \right)
$\\
explicit $X$ & $X^{M}$ & $V^{M}=\frac{1}{2}G^{MN}\partial_{N}W$\\
gamma matrix, vielbein~ & $\Gamma_{M}$ & $E_{M}^{a}\left(  X\right)
\Gamma_{a}$\\
spin connection~ & $\Gamma^{M}\partial_{M}\Psi~$ & $E^{Mc}\Gamma_{c}\left(
\partial_{M}+\frac{1}{4}\Gamma_{ab}~\omega_{M}^{ab}\left(  X\right)  \right)
\Psi$\\
real scalar field $S_{i}$ & ~$-\frac{1}{2}\partial_{M}S_{i}\partial^{M}%
S_{i}~~$ & $-\frac{1}{2}G^{MN}\partial_{M}S_{i}\partial_{N}S_{i}-aS_{i}%
^{2}R\left(  G\right)  -aL(W,S_{i}^{2})$\\
dilaton ~$\Omega$ & (extra $-\frac{1}{a}$ factor)~ & $\frac{1}{2a}%
G^{MN}\partial_{M}\Omega\partial_{N}\Omega+\Omega^{2}R\left(  G\right)
+L(W,\Omega^{2})$\\\hline
\end{tabular}
\ \ \ \ \\
\text{Table-1. Matter in curved space. The dilaton is normalized with an extra
}\left(  -a\right)  ^{-1}\text{.}%
\end{gather*}
The dilaton $\Omega$ couples to Yang-Mills fields and fermions only as
follows
\begin{equation}
S\left(  A\right)  =-\frac{1}{4}\int\left(  d^{d+2}X\right)  \sqrt{G}%
\delta\left(  W\right)  ~\Omega^{\frac{2\left(  d-4\right)  }{d-2}}~Tr\left(
F_{MN}F^{MN}\right)  . \label{Aaction}%
\end{equation}%
\begin{equation}
S_{Yukawa}\left(  \Psi,S_{i},\Omega\right)  =g_{i}\int\left(  d^{d+2}X\right)
\delta\left(  W\right)  \Omega^{-\frac{d-4}{d-2}}V^{M}\left(  \bar{\Psi}%
^{L}\Gamma_{M}\Psi^{R}S_{i}+h.c.\right)  , \label{psiHphi}%
\end{equation}
The dilaton disappears in these expressions when $d+2=6.$ In addition, even
when $d+2=6,$ the dilaton can also couple to other scalars $S_{i}\left(
X\right)  $ in the potential energy $V\left(  \Omega,S\right)  $ with the only
condition that $V\left(  \Omega,S_{i}\right)  $ has length dimension $\left(
-d\right)  $ when $\dim\left(  \Omega\right)  =\dim\left(  S_{i}\right)
=-\left(  d-2\right)  /2$. This is the only place the extra field $\Phi$
appeared in flat space in the Standard Model \cite{2tstandardM}, so that field
may or may not be the dilaton\footnote{An important additional field that was
required when $d+2\neq6$ even in flat space was a \textquotedblleft
dilaton\textquotedblright, which was named $\Phi$ in \cite{2tstandardM} and
had dimension $\dim\left(  \Phi\right)  =-\frac{d-2}{2}$ like any other scalar
field $\Omega,S_{i}$. A natural as well as economical assumption (although not
necessary) is to identify the scalar field $\Phi$ that appeared in the $4+2$
dimensional Standard Model with the dilaton field $\Omega=\Phi$ that now
appears as part of the gravity triplet $\left(  W,\Omega,G^{MN}\right)
.\label{dil}$} $\Phi=\Omega?$

We now emphasize an important property of the scalars $S_{i}$ (including the
Higgs field in the Standard Model). It turns out that, for consistency with
the Sp$\left(  2,R\right)  $ conditions (\ref{sp1}-\ref{sp5}), the
\textit{quadratic} part of the Lagrangian for any real scalar $S_{i}\left(
X\right)  $ must have exactly the same structure as the one for the dilaton
field $\Omega$. So, the \textit{quadratic} part of the action for any scalar
must have the form of the dilaton action $S\left(  \Omega\right)
=S_{G}\left(  \Omega\right)  +S_{\Omega}\left(  \Omega\right)  +S_{W}\left(
\Omega\right)  $ in Eqs.(\ref{Stot}-\ref{SW}), except for substituting
$\Omega\rightarrow S_{i},$ and except for an overall normalization
constant\footnote{A complex scalar would be constructed from two real scalars
$\varphi=\left(  S_{1}+iS_{2}\right)  /\sqrt{2}.$}. This structure has been
indicated in the table above, where the piece symbolically written as
$L(W,S_{i}^{2})$ or $L(W,\Omega^{2})$ is the piece that contributes to the
action $S_{W}$ in Eq.(\ref{SW}), which appears with a $\delta^{\prime}\left(
W\right)  $ rather than $\delta\left(  W\right)  $%
\begin{equation}
\delta\left(  W\right)  L(W,S_{i}^{2})=\delta^{\prime}\left(  W\right)
\left\{  S_{i}^{2}\left(  4-\nabla^{2}W\right)  +\partial W\cdot\partial
S_{i}^{2}\right\}  .
\end{equation}
Furthermore the same special $a=\left(  d-2\right)  /8\left(  d-1\right)  $
must appear in the action of any scalar.

This last requirement is related to the underlying Sp$(2,R)$, and is most
directly understood by analyzing the consistency of the equations of motion
for the fields $G^{MN}$, $S_{i}$ and $W$ in the same footsteps as sections
(\ref{eomG}-\ref{eomW}). The Sp$\left(  2,R\right)  $ constraint is that we
must always obtain the same kinematic equations of motion, in particular
$G_{MN}=\frac{1}{2}\nabla_{M}\partial_{M}W$ in Eq.(\ref{Gshell},\ref{sp5}),
independent of the field content in the action. This is a strong condition
that demands the stated structure for the Lagrangian for any scalar field
$S_{i}.$ Of course, in flat space this is immaterial since $R\left(  G\right)
$ is zero, but it has an important physical effect on the meaning of the
gravitational constant, as perceived by observers in the shadow worlds in $d$
dimensions, as we will see below.

There remains however the freedom of an overall normalization which, for
physical reasons, must be taken as specified in the table above. Namely, for
the dilaton, the sign of the term $\Omega^{2}R\left(  G\right)  $ must be
positive since this is required by the positivity condition of gravitational
energy in the conformal shadow as seen from Eq.(\ref{shadowGrav}). Since the
dilaton is gauge freedom in the conformal shadow, the sign or normalization of
the term $\frac{1}{2a}G^{MN}\partial_{M}\Omega\partial_{N}\Omega$ was not
crucial. However, for the remaining scalar fields the sign and normalization
of the kinetic term $-\frac{1}{2}G^{MN}\partial_{M}S_{i}\partial_{N}S_{i}$
must be fixed by the requirements of unitarity (no negative norm fluctuations)
and conventional definition of norm.

It is interesting that there is a physical consequence. We consider again the
conformal shadow and try to interpret the physical structure for observers in
the smaller $d$ dimensional space. The conformal shadow is obtained by the
same steps as before by taking $W\left(  X\right)  =w$. We concentrate only on
the scalars and the metric. These fields have the following shadows%
\begin{align}
G^{MN}\left(  w,u,x\right)   &  =%
\begin{array}
[c]{cc}%
M\backslash N &
\begin{array}
[c]{ccc}%
w & \;u~\; & \;\;\;\nu\;\;\;
\end{array}
\;\;\;\\%
\begin{array}
[c]{c}%
w\\
u\\
\mu
\end{array}
& \left(
\begin{array}
[c]{ccc}%
4w & -1 & 0\\
-1 & \;\;\;0 & 0\\
0 & \;\;\;0 & e^{4u}g^{\mu\nu}\left(  x\right)
\end{array}
\right)
\end{array}
,\;\\
\Omega\left(  w,u,x\right)   &  =e^{\left(  d-2\right)  u}\phi\left(
x\right)  ,\;\;S_{i}\left(  w,u,x\right)  =e^{\left(  d-2\right)  u}%
s_{i}\left(  x\right)
\end{align}
The action in the conformal shadow at $w=0$ is then\footnote{We must
be careful that the equations of motion derived from this action are
consistent with the original equations of motion in $d+2$
dimensions. In fact, this is not trivial. The shadow extends to
$w,u$ space through first and second order terms in the expansion in
powers of $w,$ such as
\begin{align*}
g_{\mu\nu}\left(  x,we^{4u}\right)   &  =g_{\mu\nu}\left(  x\right)
+we^{4u}\tilde{g}_{\mu\nu}\left(  x\right)
+ w^2e^{8u} \tilde{\tilde{g}}_{\mu\nu}\left(  x\right)  +\cdots\\
\Omega\left(  x,we^{4u}\right)   &  =\phi\left(  x\right)
+we^{4u}\tilde {\phi}\left(  x\right) +
 w^2e^{8u} \tilde{\tilde {\phi}}\left(  x\right) +\cdots\\
S_{i}\left(  x,we^{4u}\right)   &  =s_{i}\left(  x\right)
+we^{4u}\tilde {s}_{i}\left(  x\right) +
 w^2e^{8u} \tilde{\tilde {s}}_{i}\left(  x\right) +\cdots
\end{align*}
The Riemann tensor $R_{MNPQ}\left(  G\right)  $ constructed from
$G_{MN}\left(  w,u,x\right)  $ contains the modes
$\tilde{g}_{\mu\nu}, \tilde{\tilde{g}}_{\mu\nu}  $ even after
setting $w=0$ because there are derivatives with respect to $w.$
Thus, we emphasize that $R_{\mu\nu\lambda\sigma}\left( G\right)  $
at $w=0$ depends on $g_{\mu\nu}$, $\tilde{g}_{\mu\nu}$ and
$\tilde{\tilde{g}}_{\mu\nu}$ so it is not the same as
$R_{\mu\nu\lambda\sigma}\left(  g\right) ,$ and similarly for other
components. Consistency with the full set of equations of motion
given above require also the modes $\tilde{\phi},
\tilde{\tilde{\phi}} , \tilde{s}_{i}, \tilde{\tilde{s}}_{i} $.
However, all extra modes get determined in terms of only
$g_{\mu\nu},\phi,s_{i}$ self consistently through the full set of
equations of motion in $d+2$ dimensions. The self consistent
dynamics in shadow space $x^{\mu},$ is then determined only by
$g_{\mu\nu}\left(  x\right)  ,$ and the interactions among fields
involve only $\phi\left(  x\right)  $ and $s_{i}\left( x\right)  .$
Their consistent interactions, as derived from the original
equations of motion, are then described by the shadow action given
here. These technical details will be given in a separate paper.}%
\begin{equation}
S\left(  g,\phi,s_{i}\right)  =\int d^{d}x\sqrt{-g}\left(
\begin{array}
[c]{c}%
\frac{1}{2a}g^{\mu\nu}\partial_{\mu}\phi\partial_{\nu}\phi-\frac{1}{2}%
g^{\mu\nu}\partial_{\mu}s_{i}\partial_{\nu}s_{i}\\
+\left(  \phi^{2}-as_{i}^{2}\right)  R-V\left(  \phi,s_{i}\right)
\end{array}
\right)  .
\end{equation}
Due to the special value of $a$ there is \textit{one} overall local Weyl
symmetry which can be used to fix the gauge
\begin{equation}
\phi\left(  x\right)  =\phi_{0}%
\end{equation}
as discussed above. So, $\phi\left(  x\right)  $ disappears, while the
remaining scalar fields $s_{i}\left(  x\right)  $ are correctly normalized and
are physical. The modified Einstein equation that follows from this action is%
\begin{equation}
R_{\mu\nu}-\frac{1}{2}g_{\mu\nu}R=T_{\mu\nu}\left(  \phi_{0},s_{i}\right)  ,
\end{equation}
with the energy momentum tensor given by%
\begin{equation}
T_{\mu\nu}=\frac{1}{\left(  \phi_{0}^{2}-a\sum_{i}s_{i}^{2}\right)  }\left[
\begin{array}
[c]{c}%
\sum_{i}\left(  \frac{1}{2}\partial_{\mu}s_{i}\partial_{\nu}s_{i}-\frac{1}%
{4}g_{\mu\nu}\partial s_{i}\cdot\partial s_{i}\right)  \\
-\frac{1}{2}g_{\mu\nu}V\left(  \phi_{0},s_{i}\right)  +a\sum_{i}\left(
g_{\mu\nu}\nabla^{2}s_{i}^{2}-\nabla_{\mu}\partial_{\nu}s_{i}^{2}\right)
\end{array}
\right]  .\label{Tmunu}%
\end{equation}
The trace of this energy momentum tensor is%
\begin{equation}
g^{\mu\nu}T_{\mu\nu}=\frac{\left(  d-2\right)  }{8\left(  \phi_{0}^{2}%
-a\sum_{i}s_{i}^{2}\right)  }\left[  -\frac{4d}{d-2}V\left(  \phi_{0}%
,s_{i}\right)  +2\sum_{i}s_{i}\nabla^{2}s_{i}\right]
\end{equation}
After using the equations of motion $\nabla^{2}s_{i}=\partial V/\partial
s_{i}+2as_{i}R,$ the special value of $a,$ and the homogeneity of the
potential $\left(  \phi\frac{\partial V}{\partial\phi}+\sum_{i}s_{i}%
\frac{\partial V}{\partial s_{i}}\right)  =\frac{2d}{d-2}V,$ we compare to the
trace of Eq.(\ref{Rmn}), $\left(  1-d/2\right)  R=g^{\mu\nu}T_{\mu\nu},$ and
solve for $R.$ We obtain
\begin{equation}
R\left(  g\right)  =\frac{1}{2\phi_{0}}\frac{\partial V\left(  \phi_{0}%
,s_{i}\right)  }{\partial\phi_{0}}.
\end{equation}
This is the same result as starting with the equation of motion for
$\phi\left(  x\right)  $ and then choosing the gauge $\phi\left(  x\right)
\rightarrow\phi_{0}.$ Therefore the $\phi$ equation of motion is recovered
from the equations of motion of the other fields, showing consistency.

When the $s_{i}$ are small fluctuations, $\phi_{0}^{-2}$ approximates the
overall factor in $T_{\mu\nu}$. Then the gravitational constant is determined
approximately by $\phi_{0},$ as specified in Eq.(\ref{gravConstant}).

However, if $V\left(  \phi_{0},s_{i}\right)  $ has non-trivial minima that
lead to non-trivial vacuum expectation values for some of the $\langle
s_{i}\rangle=v_{i},$ then in that vacuum the gravitational constant is
determined by
\begin{equation}
16\pi G_{d}=\left(  \phi_{0}^{2}-av_{i}^{2}\right)  ^{-1}%
\end{equation}
rather than only $\phi_{0}^{-2}.$ The massless Goldstone boson, which is
removed by the Weyl symmetry, is then a combination of $\phi$ and the scalars
$s_{i}$ that developed vacuum expectation values.

Such phase transitions of the vacuum can occur in the history of the universe
as it expands and cools down. This is represented by an effective $V\left(
\phi,s_{i}\right)  $ that changes with temperature. So, the various $v_{i}$
may turn on as a function of temperature $v_{i}\left(  T\right)  $ or
equivalently as a function of time. Among the phase transitions to be
considered is inflation, possible grand unification symmetry breaking,
electroweak symmetry breaking, as well as some possible others in the context
of string theory to determine how we end up in 4 dimensions with a string
vacuum state compatible with the Standard Model.

It would be interesting to pursue the possibility of a changing effective
gravitational constant, as above, since this cosmological scenario is now well
motivated by 2T-physics. This scenario may not have been investigated before.

\section{Comments}

As expected naively, extra timelike dimensions potentially introduce ghosts
(negative probabilities) as well as the possibility of causality violation,
leading to interpretational problems. However, 2T-physics overcomes these
problems by introducing the right set of gauge symmetries, thus correctly
describing the physical world, including the physics of the Standard Model of
particles and forces \cite{2tstandardM}\cite{SUSY2007}, and now General Relativity.

At the same time 2T-physics also gives additional physical information which
is not encoded in 1T-physics. This is because according to 2T-physics there is
a larger spacetime in $d+2$ dimensions $X^{M}$ where the fundamental rules of
physics are encoded. These rules include a complete symmetry of
position-momentum $X^{M},P_{M}$ according to the principles of a local
Sp$\left(  2,R\right)  $ with generators $Q_{ij}(X,P).$ This leads effectively
to gauge symmetries in $d+2$ dimensions that can remove degrees of freedom and
create a holographic shadow of the $d+2$ universe in $d$ dimensions $x^{\mu}$.
There are many such shadows, and since observers in different shadows use
different definitions of time, they interpret their observations as different
1T dynamics. However, the shadows are related since they represent the same
higher dimensional universe. These predicted relations would be interpreted as
dualities by observers that live in the lower dimension $x^{\mu}$ that use
1T-physics rules. With hard work, observers in the smaller $x^{\mu}$ space
could discover enough of these dualities among the shadows to reconstruct the
$d+2$ dimensional highly symmetric universe. 2T-physics provides a road map
for this reconstruction by predicting the properties of the shadows.

Examples of some simple dualities in $d$ dimensions, that arise from flat
$d+2$ dimensional spacetime, in the context of field theory such as the
Standard Model, were discussed in (\cite{emergentfieldth1}%
,\cite{emergentfieldth2}). In the flat case, each shadow has SO$\left(
d,2\right)  $ global symmetry as hidden symmetry, where this SO$\left(
d,2\right)  $ is the shadow of the global Lorentz symmetry in $d+2$ dimensions
as identified in Eq.(\ref{flatQ}). So clues of the higher spacetime can also
appear within each shadow in the form of hidden symmetries. Examples of these
in field theory were also discussed in (\cite{emergentfieldth1}%
,\cite{emergentfieldth2}).

In curved spacetime, the details of the shadow as seen by observers stuck in
the smaller spacetime $x^{\mu},$ depends partially on the choice of $W$ as a
function of $\left(  w,u,x^{\mu}\right)  .$ In this paper we discussed the
\textquotedblleft conformal shadow\textquotedblright\ defined by $W\left(
w,u,x^{\mu}\right)  =w$ in Eq.(\ref{Wchoice}) and the gauge fixed form of the
metric (\ref{GMN}). Together, these define the timeline in the shadow space
$x^{\mu}$ as some curve embedded in the 2-time spacetime in $d+2$ dimensions.
A different choice of gauges leads to a different shadow space with a
different timeline. The same dynamics in $d+2$ dimensions $X^{M}$ tracked as a
function of one timeline can appear to be quite different 1-time dynamics
relative to another timeline. Evidently, there are many choices that
correspond to many embeddings of $d$ dimensional spacetime $x^{\mu}$ (with 1
time) into $d+2$ dimensional spacetime $X^{M}$ (with 2 times), and these are
expected to lead to dualities that relate the different looking 1-time
dynamics. Depending on the nature of the higher curved space $X^{M}$, there
could be hidden symmetries that would be seen in each smaller $x^{\mu}$ space
as clues of the extra space and time.

The kinds of predictions above can be used to generate multiple tests of
2T-physics. This line of investigation is at its infancy and is worth pursuing vigorously.

In addition to the above, the emergent 1T-physics conformal shadow seems to
come with certain natural constraints, which remarkably are not in
contradiction with known phenomenology so far. On the contrary, they lead to
some new guidance for phenomenology:

\begin{itemize}
\item The Standard Model is correctly reproduced as a shadow\footnote{The
theta term $\theta F\ast F$ can be reproduced as a shadow in 3+1
dimensions from 2T field theory in 4+2 dimensions (to appear). So a
previous claim of the resolution of strong CP violation without an
axion \cite{2tstandardM} is retracted.}, but in addition, the Higgs
sector is \textit{required} to interact with an additional scalar
$\Phi$ that induces the electroweak phase transition as discussed in
\cite{2tstandardM} ($\Phi$ could be the dilaton $\Omega$, but not
necessarily, see footnote \ref{dil}). This leads to interesting
physics scenarios at LHC energy scales (an additional new neutral
scalar) or cosmological scales (inflaton candidate, dark matter
candidate) as suggested in \cite{2tstandardM}\footnote{Scenarios
that include such a scalar field in both theoretical and
phenomenological contexts have been discussed independently in
recent papers \cite{shapashnikov}-\cite{foot1} that mainly appeared
after \cite{2tstandardM}.}. The supersymmetric\footnote{It was
suggested in the second reference in \cite{2tstandardM} that a
conformal scalar of the type $\Phi,$ with the required SO$\left(
4,2\right)  $, could provide an alternative to supersymmetry as a
mechanism that could address the mass hierarchy problem. This
possibility has been more recently discussed in
\cite{nicolai2}\cite{foot2}.} version \cite{SUSY2007} of this
2T-physics feature with extra required scalars leads to richer
phenomenologically interesting possibilities.

\item The gravitational constant could be time dependent as described in the
previous section. This is because according to 2T-physics, if there are any
fundamental scalars $s_{i}\left(  x\right)  $ at all, they all must be
conformal scalars coupled to the curvature term $R$ with the special
coefficient $\left(  -a\right)  $ as in the last line line of the table above.
It would be interesting to study the effects of this scenario in the context
of cosmology.
\end{itemize}

There are many open questions. In particular quantization in the path integral
formalism is still awaiting clarification of the gauge symmetries so that
Faddeev-Popov techniques can be correctly applied. Other issues include the
question of whether there might be some physical role, either at the classical
or quantum levels, for the \textquotedblleft remainders\textquotedblright\ in
the expansion of the fields in powers of $W,$ as in Eq.(\ref{expand}).

Having accomplished a formulation of gravity as well as supersymmetry in 2T
field theory \cite{SUSY2007} it is natural to next try supergravity. In
particular the 2T generalization of 11-dimensional supergravity is quite
intriguing and worth a few speculative comments. If constructed, such a theory
will provide a low energy 2T-physics corner of M-theory. This would be a
theory in 11+2 dimensions whose global supersymmetry can only be OSp(1%
$\vert$%
64), so it should be related to S-theory \cite{Stheory}. We remind the reader
that S-theory gives an algebraic BPS-type setting based on OSp(1%
$\vert$%
64) for the usual M-theory dualities among its corners, with 11 dimensions or
10 dimensions with type IIA, IIB, heterotic, type-I supersymmetries. A
corresponding 2T-physics theory would provide a dynamical basis that could
give shadows-type meaning to these famous dualities, as outlined in
\cite{super2t}.

Finally, let us emphasize that the fundamental concept behind 2T-physics is
the momentum-position symmetry based on Sp$\left(  2,R\right)  .$ Despite the
fact that the worldline approach in Eq.(\ref{actionW}) treats position and
momentum on an equal footing, the field theoretic approach that we have
discussed blurs this symmetry, although the constraints implied by the
Sp$\left(  2,R\right)  $ symmetry in the form of the kinematic constraints
were still maintained. There should be a more fundamental approach with a more
manifest position-momentum symmetry, perhaps with fields that depend both on
$X^{M}$ and $P_{M}$, and in that case perhaps based on non-commutative field
theory. Basic progress along this line that included fields of all integer
spins was reported in \cite{2tfieldXP}. If this avenue could be developed to a
comparable level as the current field theory formalism, it is likely that it
will go a lot farther than our current approach.

\begin{acknowledgments}
I would like to thank S.-H Chen, Y.-C. Kuo and B. Orcal for useful
discussions. Special thanks to E. Witten for his support of a visit to the IAS
where the last stages of this work was completed, and for encouraging me to
pursue more vigorously the phase space approach outlined in the last
paragraph. I also thank N. Arkani-Hamed and J. Maldacena for helpful
discussions on conformal scalars and anomalies.
\end{acknowledgments}

\end{document}